\documentclass[pra,showpacs,bibnotes,twocolumn]{revtex4}
\usepackage{float}
\usepackage{graphicx}
\begin{document}
\draft
\def\ds{\displaystyle}
\title{$\mathcal{P}\mathcal{T}$ Symmetric Aubry-Andre Model}
\author{ C. Yuce}
\address{Department of Physics, Anadolu University,
Eskisehir, Turkey.\\ cyuce@anadolu.edu.tr}
\email{cyuce@anadolu.edu.tr}
\date{\today}

\begin{abstract}
$\mathcal{P}\mathcal{T}$ Symmetric Aubry-Andre Model describes an
array of $\ds{N}$ coupled optical waveguides with position
dependent gain and loss. We show that the reality of the spectrum
depends sensitively on the degree of disorder for small number of
lattice sites. We obtain the Hofstadter Butterfly spectrum and
discuss the existence of the phase transition from extended to
localized states. We show that rapidly changing periodical
gain/loss materials almost conserves the total intensity.
\end{abstract}
\maketitle

\section{Introduction}

Recent experimental realization of $\mathcal{P}\mathcal{T}$
symmetric optical systems with balanced gain and loss has
attracted a lot of attention \cite{deney1,deney2,deney3}. The
$\mathcal{P}\mathcal{T}$ symmetric optical systems lead to
interesting results such as unconventional beam refraction and
power oscillation \cite{powerosc,refr,ek2}, nonreciprocal Bloch
oscillations \cite{bloch}, unidirectional invisibility \cite{inv},
an additional type of Fano resonance \cite{fano}, and chaos
\cite{chaos}. In the $\mathcal{P}\mathcal{T}$ symmetric optical
systems, the net gain or loss of particles vanishes due to the
balanced gain and loss mechanism. These systems are described by
non-Hermitian Hamiltonian with real energy eigenvalues provided
that non-Hermitian degree is below than a critical number,
$\ds{\gamma_{PT}}$. If it is beyond the critical number,
spontaneous $\mathcal{P}\mathcal{T}$ symmetry breaking occurs. It
implies the eigenfunctions of the Hamiltonian are no longer
simultaneous eigenfunction of $\mathcal{P}\mathcal{T}$ operator
and consequently the energy spectrum becomes either partially or
completely complex. The critical number of non-Hermitian degree is
shown to be different for planar and circular array configurations
\cite{ek} and it can be increased if impurities and tunneling
energy are made position-dependent in an extended lattice
\cite{jog1}. However, $\ds{\gamma_{PT}}$ decreases with increasing
the lattice sites \cite{bendix2,bendix,jog0,jog2}, hence the
$\mathcal{P}\mathcal{T}$ symmetric phase is fragile. An important
consequence of $\mathcal{P}\mathcal{T}$ symmetric optical systems
is the power oscillations. It was shown that the beam power in a
one dimensional tight binding chain doesn't depend on the
microscopic details such as disorder and periodicity
\cite{random2}. The probability-preserving time evolution in terms
of the Dirac inner product for $\mathcal{P}\mathcal{T}$ symmetric
tight-binding ring was considered \cite{probConserv}. It is
interesting to note that the $\mathcal{P}\mathcal{T}$ operator
coincides with time evolution operator at some certain times that
allows perfect state transfer in the $\mathcal{P}\mathcal{T}$
symmetric optical lattice with position dependent tunneling energy
\cite{stateTrans}. The equivalent Hermitian Hamiltonian for a
tight-binding chain can also be constructed to understand the
non-Hermitian system \cite{equHerm}.\\
In this paper, we investigate disordered array of
$\mathcal{P}\mathcal{T}$ symmetric tight binding chain
\cite{random1,disorder1,disorder2,anderson}. It is well known that
disorder in quantum mechanical systems induces localization. Here,
we show that localization occurs if some certain conditions are
satisfied. Our system is described by the $\mathcal{P}\mathcal{T}$
symmetric extension of the Aubry Andre model \cite{aubry}. The
energy spectrum associated with Hermitian Aubry-Andre model at
certain strength of parameters has fractal structure, which is
known as the Hofstadter butterfly spectrum \cite{HB,harper}. We
also investigate the Hofstadter butterfly spectrum in the presence
of non-Hermitian impurities.

\section{Model}

\begin{figure}[t]
\begin{center}
\centering
\includegraphics[width=5cm]{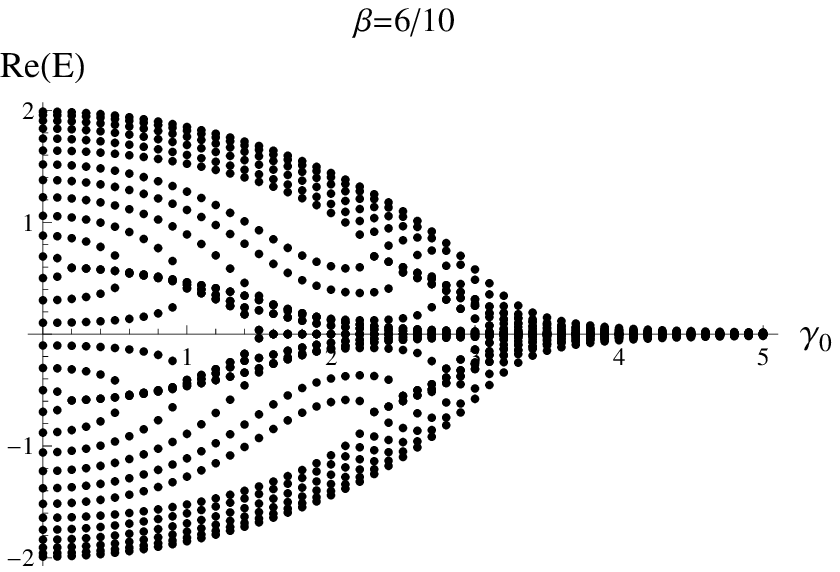}
\includegraphics[width=5cm]{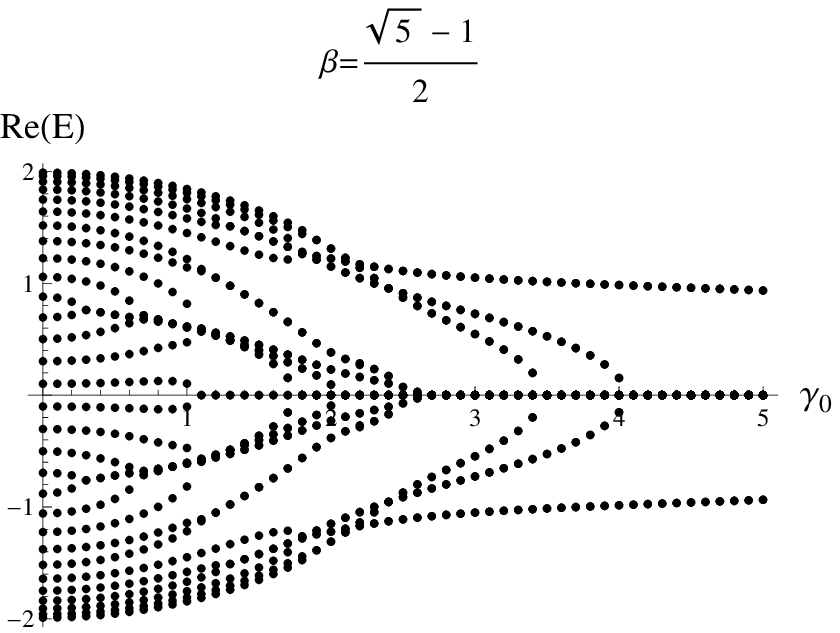}
\caption{The real part of energy eigenvalues as a function of
$\gamma_0$ at $N=30$, $V=0$ and $J=1$. We take $\ds{\beta=0.6}$
and $\ds{\beta=(\sqrt{5}-1)/2\approx0.618}$.} \label{cuwe}
\end{center}
\end{figure}

\begin{figure}[t]
\begin{center}
\centering
\includegraphics[width=4.25cm]{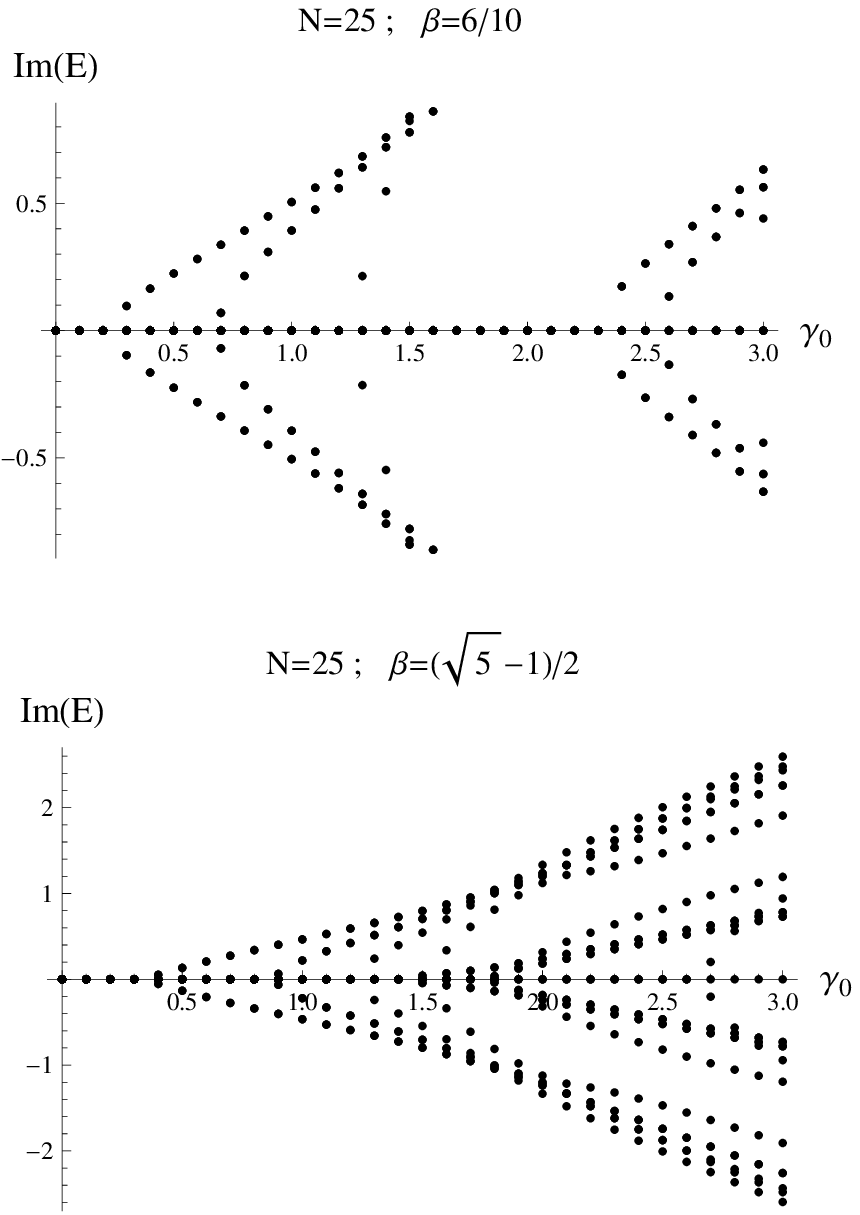}
\includegraphics[width=4.25cm]{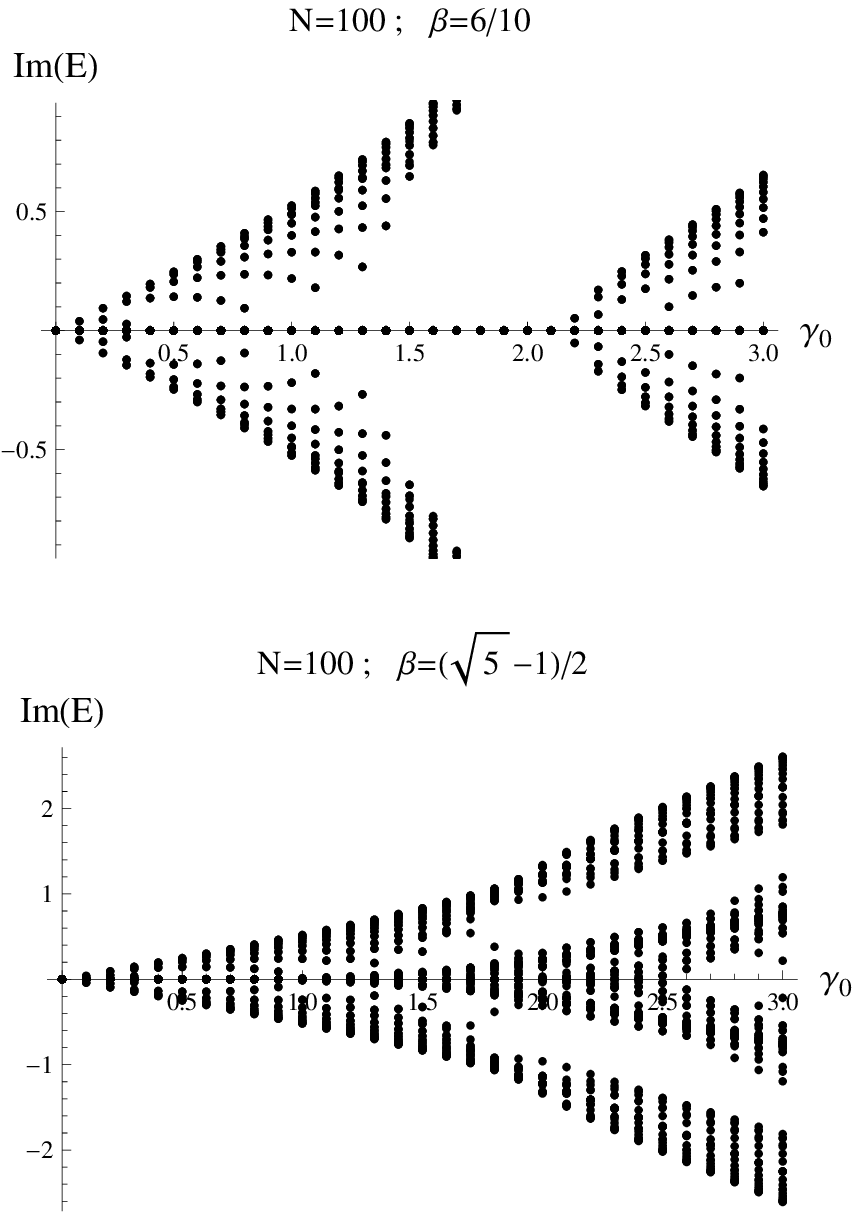}
\caption{The imaginary part of energy eigenvalues as a function of
$\gamma_0$ at $J=1$, $V=0$ and $N=10,100$. The degree of the
disorder changes $\gamma_{PT}$ significantly. $\gamma_{PT}$ for
$\ds{\beta=(\sqrt{5}-1)/2}$ is nearly $1.6$ times greater than
$\gamma_{PT}$ for $\ds{\beta=6/10}$ when $N=25$. For large number
of lattice sites, $\gamma_{PT}\rightarrow0$.} \label{gamma}
\end{center}
\end{figure}

Consider an array of $\ds{N}$ coupled optical waveguides with
position dependent gain and loss and constant tunneling amplitude
$\ds{J}$ through which light is transferred from site to site. We
adopt open boundary conditions. The beam propagation in the
tight-binding structure can be described by a set of equations for
the electric field amplitudes $c_n$,
\begin{eqnarray}\label{denklem0}
i\frac{d{c}_{n}}{dz}=-J\left(c_{n+1}+c_{n-1}\right)+i\gamma_nc_{n}~,
\end{eqnarray}
where $n=1,2,...,N$ is the waveguide number and the position
dependent non-Hermitian degree $\ds{\gamma_n}$ describes the
strength of gain/loss material that is assumed to be balanced,
i.e., $\ds{\sum_{n=1}^{N}\gamma_n=0}$. The field amplitude
transforms as $\ds{c_{n}{\rightarrow}c_{N-n+1}}$ under parity
transformation and the complex number transforms as
$\ds{i{\rightarrow}-i}$ under anti-linear time reversal
transformation. Thus the global $\mathcal{P}\mathcal{T}$ symmetry
is lost unless a precise relation between $\ds{\gamma_n}$ holds.
To model disorder, $\gamma_N$ can be chosen randomly with zero
mean \cite{random2}. In this case, the system would be no longer
$\mathcal{P}\mathcal{T}$ symmetric and the corresponding energy
eigenvalues are not real. Bendix et al. studied a disordered
system by considering a pair of $N$ coupled dimers with impurities
$(\gamma_n,-\gamma_n)$ \cite{bendix}. They noted that the system
is not $\mathcal{P}\mathcal{T}$ symmetric as a whole (global
symmetry), but it possesses a local $\mathcal{P}_d\mathcal{T}$
symmetry that admits real spectrum. Here, we consider a disordered
system with global $\mathcal{P}\mathcal{T}$ symmetry and study
localization, which is well known to occur in a disordered
Hermitian lattice. Consider the following gain/loss parameter
\begin{equation}\label{gamman}
i\gamma_n=V\cos\left(2\pi\beta{n}+\phi_N\right)+i\gamma_0\sin\left(2\pi\beta{n}+\phi_N\right),
\end{equation}
where $V$ and $\gamma_0$ are constants, $\ds{\beta}$ determines
the degree of the disorder and $\phi_N$ is the constant phase
difference which depends on the total number of sites. We require
that gain and loss are balanced, so we demand
$\ds{\phi_N=-\pi\beta(N+1)+\phi_0}$, where the constant
$\ds{\phi_0}$ is an integer multiple of $\ds{\pi}$. Without loss
of generality, we take $\phi_0=0$. We emphasize that the system is
$\mathcal{P}\mathcal{T}$ symmetric globally. The Equ.
(\ref{denklem0}) with (\ref{gamman}) can be called the
$\mathcal{P}\mathcal{T}$ symmetric Aubry-Andre model \cite{aubry},
which can now be engineered experimentally
\cite{deney1,deney2,deney3}. The most interesting result of the
Hermitian Aubry-Andre model is that the states at the center of
the lattice is localized (Anderson localization) for irrational
values of $\beta$ when $\ds{V>2}$. Apparently, the non-Hermitian
character of the Aubry-Andre equation (\ref{denklem0}) could
change the physics of this system dramatically. Note also that
Aubry-Andre model coincides with the Harper model when $V=2$ and
$\gamma_0=0$ and the energy spectrum as a function of $\beta$ is
known as Hofstadter butterfly spectrum, which is an example of
fractal structure that appears in physics \cite{HB,harper}. Here,
we study the Hofstadter butterfly spectrum and localization effect
for the $\mathcal{P}\mathcal{T}$
symmetric Aubry-Andre model.\\
It is sufficient to analyze the region $\ds{0<\beta<1}$ since the
system repeats itself in equal intervals of $\ds{\beta}$.
Furthermore, the energy spectrum is symmetric with respect to
$\ds{\beta=0.5}$ axis and the spectrum does not depend on the sign
of $\gamma_0$. As a special case, if $\beta$ is either $0$ or $1$,
then the gain/loss terms vanish. If $\beta=1/2$ and $N$ is even,
the system has gain and loss with amplitudes ${\mp}i\gamma_0$ at
alternating lattice sites. The gain/loss materials change
periodically if $\beta$ is a rational number and
quasi-periodically if $\beta$ is an irrational number. In the
latter case, the gain/loss impurities are disordered. Note that
$\ds{\beta}$ can be given with a finite number of digits in a real
experiment. To increase the incommensurability of $\ds{\beta=p/q}$
($p$, $q$ are two coprime positive integers), one can choose
sufficiently large $p$ and $q$. Then the system becomes strongly
disordered.\\
\begin{figure}[t]
\begin{center}
\centering
\includegraphics[width=5.5cm]{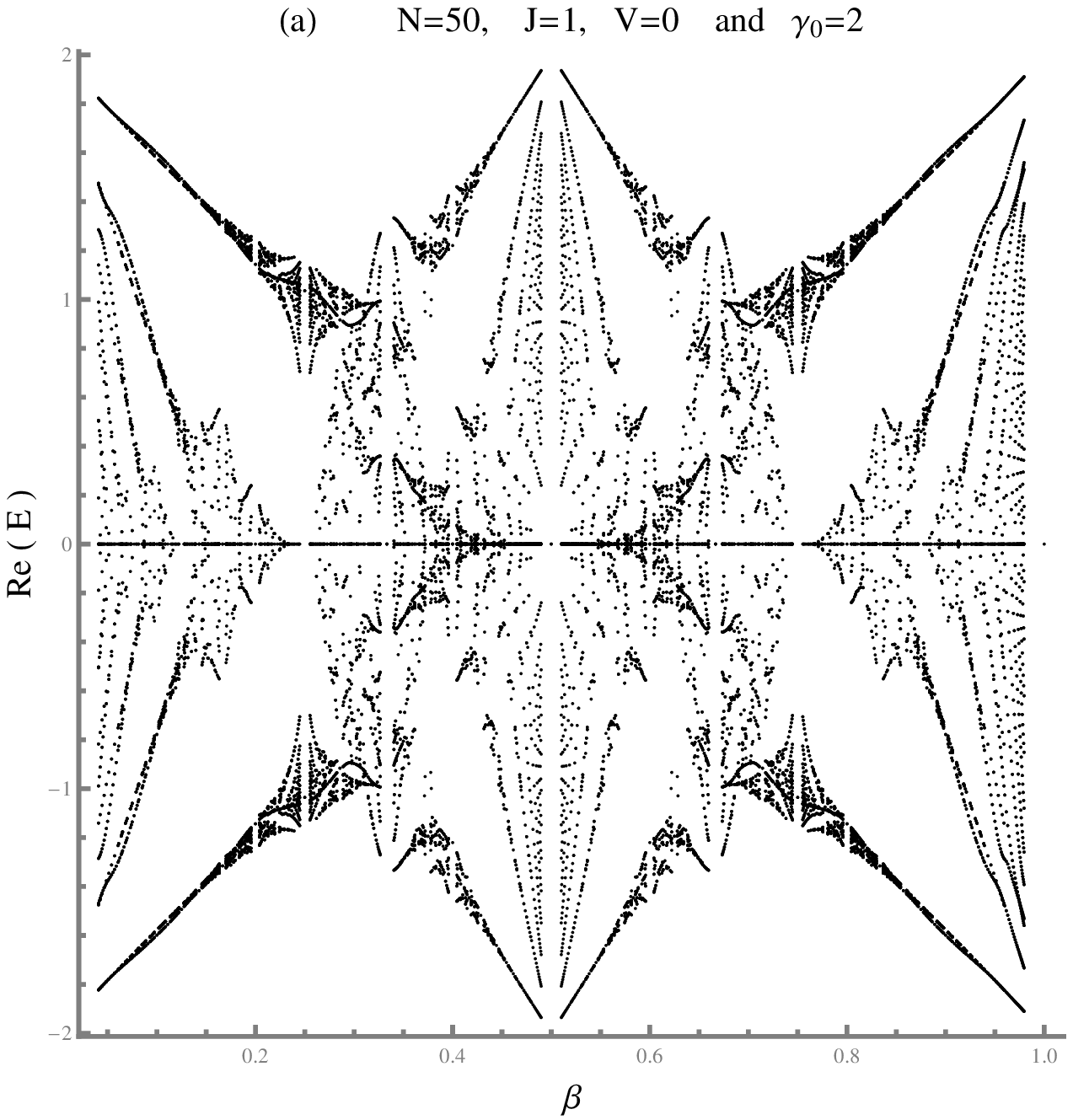}
\includegraphics[width=5.5cm]{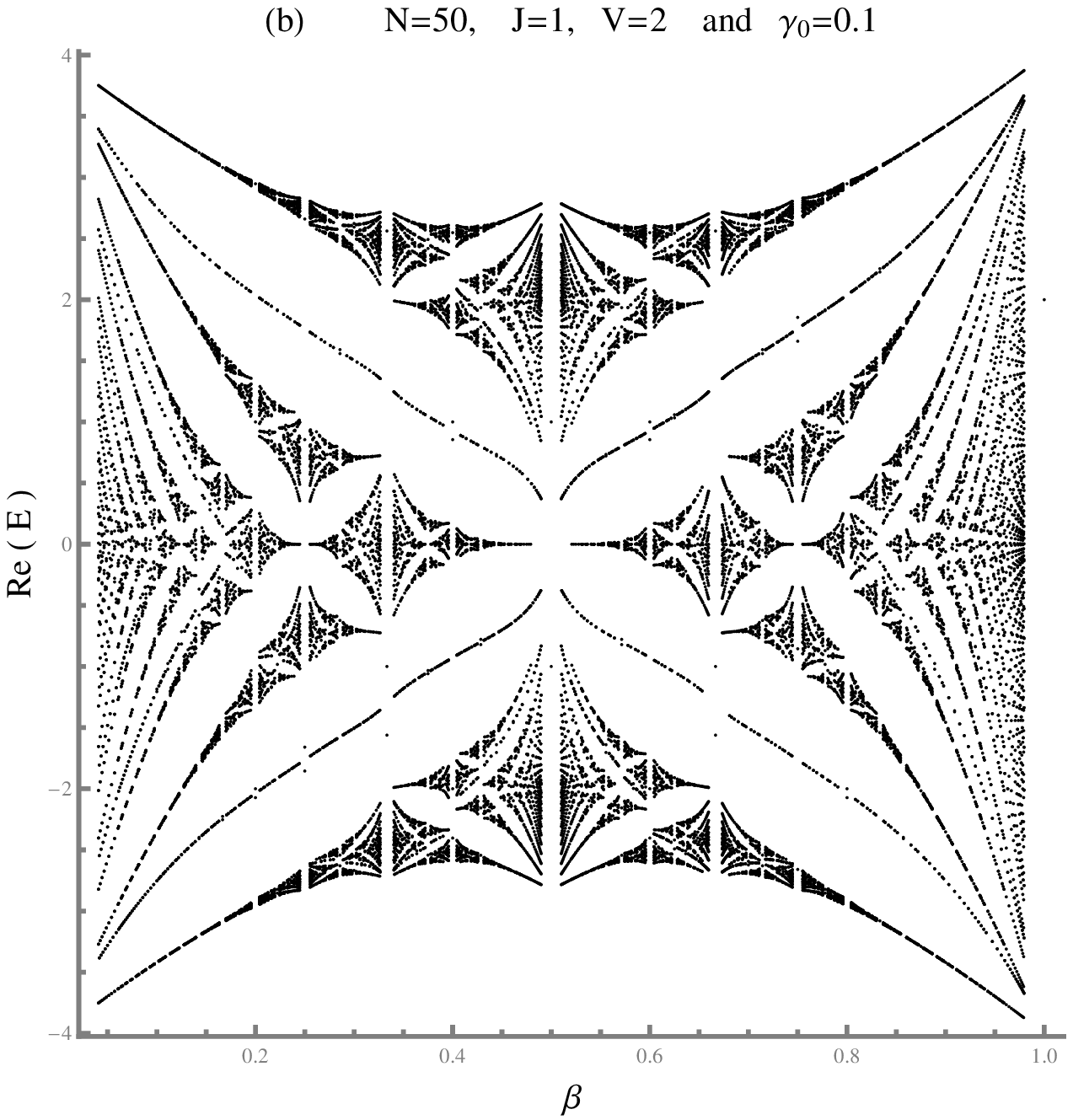}
\caption{ The $\mathcal{P}\mathcal{T}$ symmetric Hofstadter
butterfly spectrum for the real part of the energy eigenvalues at
$\ds{N=50}$ and $\ds{\gamma_0=2}$ (a), $\ds{\gamma_0=0.1}$ (b).
The graph has a line of reflection at $\beta=1/2$. } \label{hofs}
\end{center}
\end{figure}
We look for stationary solutions of the equation (\ref{denklem0}).
Suppose first that $\ds{V=0}$. In the absence of gain and loss,
the system has the well known energy spectrum of width $4J$:
$\ds{E=-2J \cos{n\pi/N}}$. In the presence of gain and loss, the
real part of the energy eigenvalues, $\ds{\mathcal{R}\{E\}}$, are
still contained in $\ds{[-2J,2J]}$ for any $N$. More precisely,
the energy width is a decreasing function $\ds{|\gamma_0|}$. The
distribution of $\ds{\mathcal{R}\{E\}}$ crucially depends on the
strength of disorder through the value of $\beta$. It consists of
a finite number of bands when $\beta$ is rational. In this case,
$\ds{\mathcal{R}\{E\}}$ is the union of bands and the length of
the gap between any two bands depends on $q$ $(\beta=p/q)$. On the
other hand, the fractal structure appears and the spectrum is a
Cantor set when $\beta$ is irrational (for the mathematicians,
this property is known as the 10 Martinis conjecture \cite{simon}
in the Hermitian limit). Such a fractal structure can be seen in
the Fig-(\ref{hofs}), where we plot the $\mathcal{P}\mathcal{T}$
symmetric Hofstadter butterfly spectrum at $\ds{\gamma_0=2}$,
$\ds{V=0}$ (a) and at $\ds{V=2}$, $\ds{\gamma_0=0.1}$ (b). An
important difference between $\ds{V{=}0}$ and $\ds{V{\neq}0}$
cases is that the symmetry with respect to zero energy axis is
lost for the latter case. However, the real part of the energy
eigenvalues is symmetric with respect to $\beta=0.5$ axis for any
$V$. Note also that the width of $\ds{\mathcal{R}\{E\}}$ increases
with $V$ and it takes its maximum value when $\gamma_0=0$. We show
the nice fractal picture for the real part of the spectrum. As can
be seen below, the $\mathcal{P}\mathcal{T}$ symmetry breaking
point is very small for large $N$ and thus the corresponding
energy spectrum is not real. However, there exists some special
values of $\ds{\beta}$ for the Fig-(\ref{hofs}.b) with entirely
real spectrum. For example, the spectrum is real when
$\ds{\beta=1/5}$. To gain more insight on the role of disorder,
let us study how the real and imaginary parts of the energy change
with $\gamma_0$ for weakly and strongly disordered system. The
Fig-(\ref{cuwe}) plots $\ds{\mathcal{R}\{E\}}$ as a function
$\gamma_0$ for a weak $\beta=1/3$ and strong disorders
$\beta=11/30$ when $N=30$. As can be seen from the figures, the
degree of disorder in the lattice has a dramatic effect for large
values of $\gamma_0$. The real part of energy shrink to zero (they
become degenerate and the energy width becomes zero) for very
large values of $\gamma_0$ if the system is periodical while this
is not the case if it is quasi-periodical. However, the
corresponding imaginary part of energy eigenvalues are different
from zero for such a large number of $\ds{\gamma_0}$. If the
impurity strength, $\ds{\gamma_0}$, exceeds a critical point,
$\ds{\gamma_{PT}}$, $\mathcal{P}\mathcal{T}$ symmetry is
spontaneously broken and thus the energy eigenvectors are not
simultaneous eigenvectors of the Hamilton and
$\mathcal{P}\mathcal{T}$ operators. In this case, the energy
eigenvalues become partially or entirely complex. An important
consequence for our system is that strong disorder increases the
critical point $\ds{\gamma_{PT}}$ considerably. The
Fig-(\ref{gamma}) plots the imaginary part of the energy
eigenvalues for various values of $N$ at $\beta=0.6$ and the
inverse of the golden ratio $\beta=(\sqrt{5}-1)/2{\approx}0.618$,
which is the common choice in the study of the Aubry-Andre model.
We numerically find that due to the disorder, $\ds{\gamma_{PT}}$
increases by a factor of nearly $1.6$ when $N=25$. However, the
number of lattice sites $N$ has dominant effect on
$\ds{\gamma_{PT}}$ and thus increasing the degree of the disorder
has slightly changes $\ds{\gamma_{PT}}$ for large $N$. The
critical point decreases with increasing $N$ and approaches zero
when $N$ is large. The $\mathcal{P}\mathcal{T}$ symmetric phase is
said to be fragile since
$\gamma_{PT}$ is zero as $\ds{N{\rightarrow}\infty}$ \cite{bendix2}. \\
It is well known that the Hermitian Aubry-Andre model,
$\ds{\gamma_0=0}$, displays a phase transition from extended to
exponentially localized states (It is also known as metal
insulator transition \cite{metins}). Of particular importance is
the self-dual point $V/J = 2$ where the localization transition
occurs \cite{aubry}. Let us now study whether localization takes
place for the $\mathcal{P}\mathcal{T}$ symmetric Aubry-Andre
model. Suppose first $\ds{V=0}$. We take the inverse of the golden
ratio $\ds{\beta=\frac{\sqrt{5}-1}{2}}$, $J=1$ and $\ds{N=49}$
with the initial condition $\ds{|c_n(z=0)|^2=\delta_{n,25}}$. We
find numerically that initially localized wave packet delocalizes
in time when $\gamma_0=2$. We repeat numerical solution for large
values of $\gamma_0$, but exponentially localized states do not
emerge for the $\mathcal{P}\mathcal{T}$ symmetric Aubry-Andre
model. This result is interesting that disorder does not induce
localization for the $\mathcal{P}\mathcal{T}$ symmetric Aubry
Andre model contrary to Hermitian one. This is because of the
fragile nature of the $\mathcal{P}\mathcal{T}$ symmetric phase.
Before metal insulator phase transition takes place, the system
enters broken $\mathcal{P}\mathcal{T}$ symmetric phase and the
corresponding intensity grows exponentially, where the intensity
is given by $\ds{I=\sum_{n=1}^N |c_n|^2}$ and satisfies
\begin{equation}\label{power}
\frac{dI}{dt}=2 \sum_{n=1}^N \mathcal{R}\{\gamma_n\} |c_n|^2~.
\end{equation}
The intensity grows exponentially in the broken
$\mathcal{P}\mathcal{T}$ symmetric case while it oscillates when
the energy spectrum is entirely real. \\
Suppose now $\ds{V{\neq}0}$. We find numerically the time
evolution of the single site excitation. It is well known that the
metal-insulator transition occurs at $\ds{V=2}$ and
$\ds{\gamma_0=0}$ and the wave packet is localized around the
single site when $V>2$. If $V<2$, the probability $|c_{25}|^2$
goes to zero rapidly with $z$. The presence of gain/loss change
the dynamics significantly. Although the probability $|c_{25}|^2$
doesn't rapidly go to zero when $\ds{\gamma_0{\neq}0}$, this can
not be considered localization in the rigorous sense. This is
because the introduction of gain/loss to the system does not
conserve the total intensity and the generated particles enter the
system not only $n=25$-th lattice site but also the other lattice
sites. Thus the occupation on the waveguide away from $n=25$ is
not negligible. For large $z$, the particles are generated even at
the edges of the system.\\
A question arises. Does the phase transition from extended to
exponentially localized state occurs if we somehow find a way to
make the total intensity bounded for large values of
$\ds{\gamma_0}$? To answer this question, consider $z$-dependent
periodic impurity strength
\cite{time1,time2,time3,time4,time5,time6}
\begin{equation}\label{gamman3}
i\gamma_n=V\cos\left(2\pi\beta{n}+\phi_N\right)+i\gamma_0\cos({2\pi\omega}z)\sin\left(2\pi\beta{n}+\phi_N\right),
\end{equation}
where $\omega$ is a constant. Note that the corresponding
Hamiltonian is still $\mathcal{P}\mathcal{T}$ invariant. The gain
and loss are also locally balanced after one period.\\
\begin{figure}[t]
\begin{center}
\centering
\includegraphics[width=5cm]{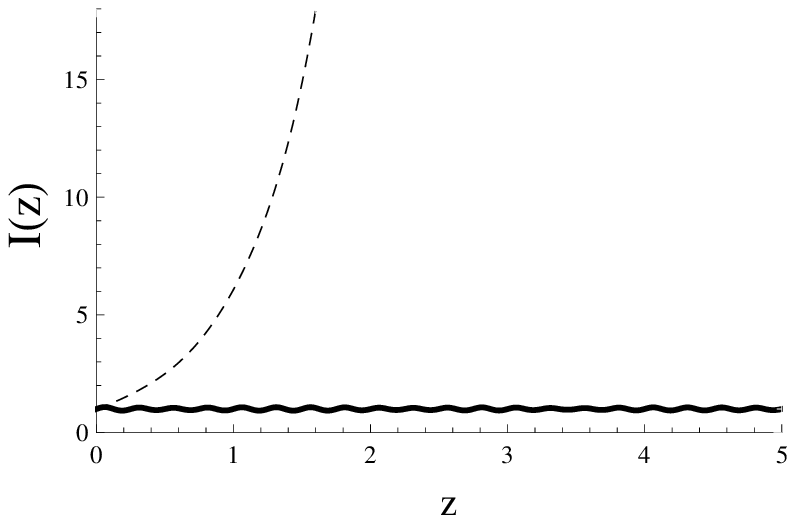}
\includegraphics[width=5cm]{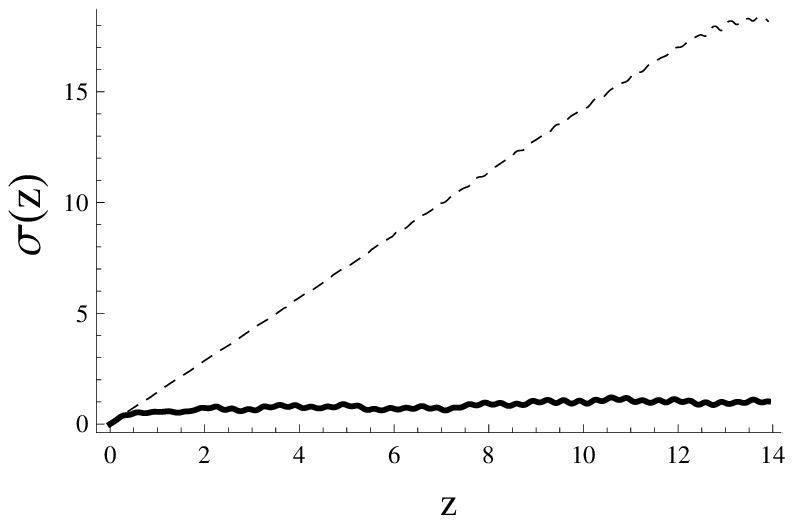}
\caption{The total intensity as a function of $z$ for the
parameters $\ds{\omega=0}$ (dashed) and $\ds{\omega=3}$ (solid) at
fixed $\ds{V=4}$. It grows exponentially for the static case while
it is almost constant for large values of $\omega$. We plot
$\ds{\sigma(z)}$ for $\ds{V=4}$ (solid) and $\ds{V=0}$ (dashed) at
fixed $\ds{\omega=3}$. Localization takes place when
$\ds{\sigma(z)}$ oscillates (non-periodically) and ballistic
expansion occurs when $\ds{\sigma(z)}$ increases linearly. We take
, $\ds{\beta=\frac{\sqrt{5}-1}{2}}$, $N=49$, $J=1$, and
$\ds{\gamma_0=2}$ for both plots. We assume that only $n=25$-th
well is occupied initially. } \label{powerbeta}
\end{center}
\end{figure}
The intensity oscillates in time when $\omega=0$ if
$\ds{\gamma_0<\gamma_{PT}}$. The oscillation is not in general
periodic. Introducing periodically changing impurity,
$\ds{\omega{\neq}0}$, makes the oscillation periodical with $z$.
Increasing $\omega$ decreases the period of the intensity. We
assert that the intensity is in principle conserved in the limit
$\ds{\omega{\rightarrow}\infty}$ since impurities do not have
enough time to transfer intensity to the system. So, we expect
that rapidly changing impurities practically conserves the
intensity. To check this argument, we solve the equation
(\ref{denklem0}) numerically. We find that the intensity is almost
constant when $\omega=3$ as can be seen from the
Fig-(\ref{powerbeta}). The disorder has nothing to do with the
intensity conservation and the intensity is
almost conserved for any values of $\beta$. \\
To predict localization, let us define the variance of the
probability distribution as $\ds{\sigma(z)=\sqrt{\sum_n
(n-\bar{n})^2 |c_n|^2/P}}$, where $\ds{\bar{n}=\sum_n n|c_n|^2/P}$
is the average site $z$-dependent occupation \cite{time6}. We plot
the variance in the Fig-(\ref{powerbeta}). The linearly increasing
$\ds{\sigma(z)}$ with respect to $z$ implies that the wave packet
delocalizes (spreads ballistically with $z$). On the contrary,
oscillating $\ds{\sigma(z)}$ shows us that the wave packet is
localized. We find that the onset of localization appears for
$\mathcal{P}\mathcal{T}$ symmetric Aubry Andre model provided that
$\ds{V/J>2}$ and the system is disordered. We emphasize that
localization doesn't take place if the system is ordered, i.e.
$\ds{\beta}$ is a rational number. In the localization regime, the
occupation at $n=25$-the well oscillates periodically with $z$ and
is almost constant for large values of $V$. We note that the
underlying mechanism of localization studied here is essentially
the same of Anderson localization.\\
To summarize, we have studied $\mathcal{P}\mathcal{T}$ symmetric
tight binding optical lattice with disordered impurities. We have
considered the complex extension Aubry-Andre model. We have
plotted complex Hofstatder butterfly spectrum and shown that the
reality of the spectrum depends sensitively on the impurity
strength and $\ds{\beta}$. We have shown that the critical point
$\gamma_{PT}$ increases with the increasing degree of disorder. We
have demonstrated that the transition from extended to localized
states does not occur for the system described by
$\mathcal{P}\mathcal{T}$ symmetric Aubry Andre model. The metal
insulator transition occurs if the impurities changes periodically
with $z$ at each site. We have also shown that rapidly changing
periodical impurities conserves the total intensity.

\end{document}